# Family of even/odd CV states, their properties and deterministic generation of the hybrid entangled states

Evgeny V. Mikheev[1] and Sergey A. Podoshvedov[2]

[1]Department of physics of nanoscale systems, Institute of natural and exact sciences, South Ural State University (SUSU), Lenin Av. 76, Chelyabinsk, Russia

[2]Laboratory of Quantum Information Processing and Quantum Computing, Institute of Natural and Exact Sciences, South Ural State University (SUSU), Lenin Av. 76, Chelyabinsk, Russia

e-mail:Evm72011@gmail.com[1]
email: sapodo68@gmail.com[2]

We consider a family of continuous variable (CV) states being a superposition of displaced number states with equal modulo but opposite in sign displacement amplitudes. Either an even or odd CV state is mixed with a delocalized photon at a beam splitter with arbitrary transmittance and reflectance coefficients with the subsequent registration of the measurement outcome in an auxiliary mode to deterministically generate hybrid entanglement. We show that at certain values of the experimental parameters maximally entangled states are generated. The considered approach is also applicable to truncated finite versions of even/odd CV states. We study the nonclassical properties of the introduced states and show their Wigner functions exhibit properties inherent to nonclassical states. Other nonclassical properties of the states under consideration have also been studied.

## I. INTRODUCTION

Entanglement, namely the property of two or even more physical systems to be described by one wave function (one state), despite the fact that these physical systems can be at a considerable distance from each other [1-3], plays a key role in the creation of a quantum computer and the implementation of quantum protocols. Therefore, the perfect correlations imposed by entanglement arouse genuine interest. Although initially Einstein, Podolsky and Rosen came to the conclusion that the quantum effect may travel faster than light, which is contradiction to the theory of relativity [4]. The correlations (EPR state) were proposed to introduce conceptions of locality and reality to quantum mechanics [4], which as it turned out to be exact opposite to any quantum theory. Later, in response to the EPR paradox, Bell proposed an inequality that could be used to test the fundamental concepts of our outlook [5]. The introduced inequalities are performed provided that we accept local realism assumption. Otherwise, it is worth considering local realism concept (either locality or realism or both of them) to be untenable, despite the fact that it does not contradict human perception. Experimental violation of Bell's inequalities was repeatedly confirmed by some quantum systems [6-9]. Large spatial separation and efficient spin read-out allows one to close the detection loophole and ensures required locality conditions [10]. The measurement settings were controlled by photons coming from distant quasars to avoid "freedom-of-choice" loophole in testing quantum mechanics [11]. In all the cases considered, local realism and quantum mechanics show completely opposite predictions. Nevertheless, EPR paradox has become the basis to define the entanglement term used for description of correlations which have no classic counterparts.

Wide practical applications of entanglement follow from development of the quantum information processing. Entanglement became the basis for quantum teleportation [12-18], quantum state engineering [19,20] and quantum computing [21-24]. Now, spontaneous parametric down-conversion (SPDC) is the standard source of entangled-photon pairs [25,26]. Introduction of another type of the entangled state is based on thought paradox [27] which was initially proposed to demonstrate weakness of the Copenhagen interpretation of quantum mechanics regarding large objects. Now optical version of the Schrödinger cat states (SCSs) is well studied [20], although the practical implementation of the states with an amplitude≥ 2 is unattainable due to the weakness of cubic nonlinearity. The hybrid entangled states is another type of entanglement formed by objects of various physical nature [28-30]. The potential of such states for quantum information processing is quite high [17,18,31,32].

Despite progress in the implementation of the entangled states no effective methods for distribution of the entanglement between distant points of a quantum network are present. Delivering the entanglement, especially in a deterministic fashion, could provide significant facilities for secure long-distance communications and powerful quantum computing. We present a method that can be used as a basis for creating multipartite entanglement. Fro the purpose, we consider the implementation of deterministic entanglement between two physical systems CV state and delocalized photon. A hybrid entangled state is generated every time a measurement outcome is detected in an auxiliary mode after the CV state is mixed with the delocalized photon on a beam splitter. The amount of entanglement (in our case, negativity [33,34]) varies in a wide range but never takes zero values. Maximum entanglement with a high success probability is observed for a large number of experimental parameters. As the CV states, we choose a family of superpositions of displaced Fock states with equal modulus but opposite in sign displacement amplitudes (generalization of the SCSs). Depending on the parity of the Fock states forming the CV states, they are divided into even and odd. The superpositions of the even/odd CV states are used for deterministic generation of entangled states and can also be used to distribute entanglement between different destinations. The method is also applicable to truncated versions of even/odd CV states, which greatly increases the possibility of its practical implementation. A combination of the approach developed with other tools can be useful for implementation of functional quantum networks. The nonclassical properties of the even/odd CV states are considered, in particular, we demonstrate their Wigner functions that demonstrate properties inherent nonclassical states.

## II. FAMILY OF EVEN/ODD CV STATES

Let us consider a family of the CV states being superposition of the displaced number states (DNSs) [35,36] whose displacement amplitudes are equal in magnitude but opposite in sign

$$|\Omega_{\pm}^{(l)}\rangle_1 = N_{\pm}^{(l)}(\beta)(|l,-\beta\rangle_1 \pm (-1)^l|l,\beta\rangle_1), \qquad (1)$$

where the DNSs $|l,\pm\beta\rangle$ with the displacement amplitudes $\pm\beta$ are determined in Appendix A (A1), amplitude $\beta > 0$ is assumed to be positive and

$$N_{\pm}^{(l)}(\beta) = \left(2\left(1 \pm (-1)^l F(2\beta) \cdot c_{ll}(2\beta)\right)\right)^{-1/2}, \qquad (2)$$

is the normalization factor, where the overall multiplier $F(\beta) = exp(-|\beta|^2/2)$ is introduced and the amplitudes $c_{ln}(\beta)$ are determined in Appendix A (A2). The subscript 1 indicates that the state occupies the first mode. In the case of $l = 0$, we have an optical analog of the Schrödinger cat states [27]

$$|\Omega_{\pm}^{(0)}\rangle_1 = N_{\pm}^{(0)}(\beta)(|0,-\beta\rangle_1 \pm |0,\beta\rangle_1), \qquad (3)$$

with the normalization factor $N_\pm^{(0)}(\beta) = \left(2(1 \pm F(2\beta))\right)^{-1/2}$. By analogy with SCSs, we name the states in Eq. (1) as superposition of displaced $l-$photon states (SDlPSs). For example, in the case of $l = 1$, we deal with a superposition of displaced single photon states (SDSPSs).

Depending on the parity of the Fock states forming the superpositions, the CV states in Eq. (1) can be divided into even and odd. Indeed, if we use the decomposition of the DNSs in the Fock basis (A2) and the property of the amplitudes with the opposite sign (A3), then we can rewrite the states in Eq. (1)

$$\left|\Omega_\pm^{(l)}\right\rangle = N_\pm^{(l)}(\beta)F(\beta)\sum_{n=0}^{\infty} c_{ln}(\beta)((-1)^{n-l} \pm (-1)^l)|n\rangle. \tag{4}$$

It follows from Eq. (4) all the states $\left|\Omega_+^{(l)}\right\rangle$ consist of even Fock states, while all the states $\left|\Omega_-^{(l)}\right\rangle$ involve only odd Fock states regardless of value $l$. Summarizing the Eq. (4), one can rewrite all the even SDnPSs as

$$\left|\Omega_+^{(l)}\right\rangle = (-1)^l 2 N_+^{(l)} F(\beta) \sum_{m=0}^{\infty} c_{l\,2m}(\beta)|2m\rangle, \tag{5}$$

while all the odd SDnPSs can be represented as

$$\left|\Omega_-^{(l)}\right\rangle = (-1)^{l+1} 2 N_-^{(l)} F(\beta) \sum_{m=0}^{\infty} c_{l\,2m+1}(\beta)|2m+1\rangle. \tag{6}$$

Since states $\left|\Omega_+^{(k)}\right\rangle$ and $\left|\Omega_-^{(m)}\right\rangle$ have different parities, they are orthogonal

$$\left\langle \Omega_+^{(k)} \middle| \Omega_-^{(m)} \right\rangle = 0. \tag{7}$$

As for the states of the same parity, they are not orthogonal

$$\left\langle \Omega_\pm^{(k)} \middle| \Omega_\pm^{(m)} \right\rangle = 2 N_\pm^{(k)} N_\pm^{(m)} \left(\delta_{km} \pm (-1)^m F(2\beta) \cdot c_{mk}(2\beta)\right), \tag{8}$$

where $\delta_{km}$ is Kronecker's symbol.

The choice of the CV states can be more significant. For example, as the CV state, one can choose a finite superposition of the CV states composed exclusively of even CV states $\left\{\left|\Omega_+^{(0)}\right\rangle, \left|\Omega_+^{(1)}\right\rangle, \left|\Omega_+^{(2)}\right\rangle, \ldots, \left|\Omega_+^{(l)}\right\rangle\right\}$ like

$$\left|\Omega_{in}^{(+)}\right\rangle_1 = N_l^{(+)} \sum_{k=0}^{l} b_k^{(+)} \left|\Omega_+^{(k)}\right\rangle_1, \tag{9}$$

where $b_1^{(+)}, b_2^{(+)}, \ldots, b_l^{(+)}$ are the amplitudes and $N_l^{(+)}$ is the normalization factor heeding nonorthogonality of the states $\left\{\left|\Omega_+^{(0)}\right\rangle, \left|\Omega_+^{(1)}\right\rangle, \left|\Omega_+^{(2)}\right\rangle, \ldots, \left|\Omega_+^{(l)}\right\rangle\right\}$ between each other. Another type of the CV states may be related to the choice of exclusively odd states $\left\{\left|\Omega_-^{(0)}\right\rangle, \left|\Omega_-^{(1)}\right\rangle, \left|\Omega_-^{(2)}\right\rangle, \ldots, \left|\Omega_-^{(l)}\right\rangle\right\}$ composing the final superposition

$$\left|\Omega_{in}^{(-)}\right\rangle_1 = N_l^{(-)} \sum_{k=0}^{l} b_k^{(-)} \left|\Omega_-^{(k)}\right\rangle_1, \tag{10}$$

where $b_1^{(-)}, b_2^{(-)}, \ldots, b_l^{(-)}$ are the amplitudes of the state and $N_k^{(-)}$ is the normalization factor.

In Fig. 1, we show the dependences of the Wigner function $W_\pm$ for even/odd SCSs $\left|\Omega_\pm^{(0)}\right\rangle$ (top two graphs), even/odd SDSPSs (middle two graphs) and even/odd CV superpositions $\left|\Omega_{in}^{(+)}\right\rangle = N_1^{(+)}\left(\left|\Omega_+^{(0)}\right\rangle + \left|\Omega_+^{(1)}\right\rangle\right)$, $\left|\Omega_{in}^{(-)}\right\rangle = N_1^{(-)}\left(\left|\Omega_-^{(0)}\right\rangle + \left|\Omega_-^{(1)}\right\rangle\right)$ (last two graphs) as function of the quadrature components $x_1$ and $x_2$. For all these plots, the value $\beta = 2$ is taken. As can be seen from the plots, the number of regions on the $x_1, x_2$ plane in which the Wigner function takes negative values $W_\pm < 0$, can only increase with increasing number $l$ which indicates the manifestation of nonclassicality. Interestingly, the Wigner functions for

superpositional states $|\Omega_{in}^{(\pm)}\rangle$ are similar in form to the Wigner functions of the SCSs $|\Omega_{\pm}^{(0)}\rangle$. The Wigner function of the states $|\Omega_{\pm}^{(1)}\rangle$ is already qualitatively different from those of the states $|\Omega_{\pm}^{(0)}\rangle$ and $|\Omega_{in}^{(\pm)}\rangle$.

The corresponding quadrature component distributions $P(X_1)$ and $P(X_2)$ for both even $|\Omega_{+}^{(0)}\rangle$, $|\Omega_{+}^{(1)}\rangle$ and $|\Omega_{in}^{(+)}\rangle = N_1^{(+)}\left(|\Omega_{+}^{(0)}\rangle + |\Omega_{+}^{(1)}\rangle\right)$ and odd states $|\Omega_{-}^{(0)}\rangle$, $|\Omega_{-}^{(1)}\rangle$ and $|\Omega_{in}^{(-)}\rangle = N_1^{(-)}\left(|\Omega_{-}^{(0)}\rangle + |\Omega_{-}^{(1)}\rangle\right)$ are presented in Figure 2 for $\beta = 2$. Graphs allow us to see peaks divorced for some distance in the distribution of $P(X_1)$ for all three states. The distribution $P(X_2)$ clearly shows interference features inherent for all three states.

Nonclassical properties [37] of the state under consideration follow plots in Figs. 3 and 4. So in Fig. 3, we show the dependence of precisions of the quadrature components $X_1 = a + a^+$ and $X_2 = (a + a^+)/i$, respectively, on size $\beta$ of the states $|\Omega_{\pm}^{(0)}\rangle$, $|\Omega_{\pm}^{(1)}\rangle$ and $|\Omega_{\pm}^{(0)}\rangle + |\Omega_{\pm}^{(1)}\rangle$. They obey the following relation $[X_1, X_2] = -2i$ and can be quantified by the standard derivations $\sigma_{x_1} = \sqrt{\langle X_1^2 \rangle - \langle X_1 \rangle^2}$ and $\sigma_{x_2} = \sqrt{\langle X_2^2 \rangle - \langle X_2 \rangle^2}$ satisfying the uncertainty principle $\sigma_{x_1}\sigma_{x_1} \geq 1$ [37]. They can be calculated using the distributions $P(X_1)$ and $P(X_2)$ shown in Figure 2. The case $\sigma_{x_1} = \sigma_{x_1} = 1$ corresponds to the vacuum state. If one of the standard derivations either $\sigma_{x_1}$ or $\sigma_{x_2}$ takes values less than 1, then such light becomes squeezed. In squeezed light (more precisely, in quadrature-squeezed light) fluctuations of one of the quadratures are suppressed. As can be seen from the plots in Figure 3, squeezing is detected only for the states $|\Omega_{in}^{(+)}\rangle = N_1^{(+)}\left(|\Omega_{+}^{(0)}\rangle + |\Omega_{+}^{(1)}\rangle\right)$ for small values of the amplitude $\beta < 1$ (upper left plot) and $|\Omega_{+}^{(0)}\rangle$ (upper right plot). In all other cases, only desqueezing is detected (in some cases, quite small, $\sigma_{x_i} \approx 1$, but $\sigma_{x_i} > 1$).

Another simplest parameter responsible for the statistics of photocounts is the Fano factor that is determined by relation of the dispersion to average number of counts $F = \langle \Delta m^2 \rangle / \langle m \rangle$ [37], where $\langle \Delta m^2 \rangle = \langle m^2 \rangle - \langle m \rangle^2$ and the procedure of averaging over the corresponding distribution is used. The number of pulses $m$ at the detector output (the number of photocounts) is periodically counted over a certain fixed small sampling interval T. This number fluctuates from experiment to experiment. Repeating this procedure repeatedly gives a set of numbers from which one can obtain the complete probabilistic characteristics of the discrete random quantity $m$. The parameter characterizes the difference between statistics and Poisson distribution. In the case of $F < 1$, sub-Poisson nonclassical light with sub-Poissonian statistics of photocounts is used. Figure 4 shows the dependences of the Fano factor $F$ depending on the size $\beta$ of the SDnPSs. As can be seen from the plots, the factor Fano $F$ of certain states can take values less than 1 only for some values of the amplitude $\beta$ which indicates the manifestation of their nonclassical properties of the states.

### III. GENERATION OF HYBRID ENTANGLED STATES

Now, we are interested in generating the hybrid entangled states in the most general case. To generate the optical entangled hybridity (macro-micro entanglement), a delocalized photon
$$|\varphi\rangle_{23} = a_0|01\rangle_{23} + a_1|10\rangle_{23}, \qquad (11)$$
is used which occupy simultaneously modes 2 and 3, where the amplitudes $a_0$ and $a_1$ satisfy the normalization condition $|a_0|^2 + |a_1|^2 = 1$. The input state $|\Omega_{\pm}^{(l)}\rangle$ is mixed with the delocalized photon on the beam splitter (BS) which is described by the following unitary matrix

$$BS = \begin{bmatrix} t & -r \\ r & t \end{bmatrix}, \tag{12}$$

where $t$ and $r$ are the real transmittance and reflectance coefficients, satisfying the normalization condition $t^2 + r^2 = 1$ as shown in Fig. 5.

After the mixing, the number of photons is recorded in the second auxiliary mode to conditionally generate the target state. It follows from Appendixes B-C that regardless of the number of registered photons $n$, the parity of the input CV state, the conditional hybrid entangled state will have the form

$$|\Delta_{2m}^{(l\pm)}\rangle_{13} = \mathfrak{N}_{2m}^{(l\pm)} \left( a_0 |\Psi_{2m}^{(l\pm)}\rangle_1 |1\rangle_3 + a_1 B_{2m}^{(l\pm)} |\Phi_{2m}^{(l\pm)}\rangle_1 |0\rangle_3 \right), \tag{13}$$

provided that even number of photons $n = 2m$ is detected. If an odd number of photons $n = 2m+1$ is recorded in the second auxiliary mode, then the generated hybrid entangled state becomes

$$|\Delta_{2m+1}^{(l\pm)}\rangle_{13} = \mathfrak{N}_{2m+1}^{(l\pm)} \left( a_0 |\Psi_{2m+1}^{(l\pm)}\rangle_1 |1\rangle_3 + a_1 B_{2m+1}^{(l\pm)} |\Phi_{2m+1}^{(l\pm)}\rangle_1 |0\rangle_3 \right). \tag{14}$$

Each of the introduced CV states $|\Psi_{2m}^{(l\pm)}\rangle$, $|\Psi_{2m+1}^{(l\pm)}\rangle$ and $|\Phi_{2m}^{(l\pm)}\rangle$, $|\Phi_{2m+1}^{(l\pm)}\rangle$ consists exclusively of either even or odd Fock states. The parity of the generated CV states depends on the parity of both measurement outcome in auxiliary made and the parity of the initial CV state. So, in the case of initial even state $|\Omega_+^{(l)}\rangle$ as well as in the case of an even measured outcome $n = 2m$, the CV states are even $|\Psi_{2m}^{(l+)}\rangle$ and odd $|\Phi_{2m}^{(l+)}\rangle$

$$|\Psi_{2m}^{(l+)}\rangle = L_{2m}^{(l+)} \sum_{p=0}^{l} x_{lp}^{(2m+)} |\Omega_+^{(p)}\rangle, \tag{15}$$

$$|\Phi_{2m}^{(l+)}\rangle = K_{2m}^{(l+)} \sum_{p=0}^{l+1} y_{lp}^{(2m+)} |\Omega_-^{(p)}\rangle, \tag{16}$$

where the CV states are formed from SDlPSs in Eq. (1), $x_{lp}^{(2m+)}$ and $y_{lp}^{(2m+)}$ are their amplitudes and $L_{2m}^{(l+)}$ and $K_{2m}^{(l+)}$ are the normalization factors. Analytical expressions for the amplitudes $x_{lp}^{(2m+)}$ and $y_{lp}^{(2m+)}$ are presented in Appendix B for partial case of input even/odd SCSs and in Appendix C for general case. The states, normalization factors and amplitudes are supplied by both lower and upper indices. Consider the meaning of the indices on example of the state in Eq. (15). So, the subscript for the states and normalization factors indicates the parity of the measured photons $2m$ in the auxiliary mode while superscript $(l +)$ tells that the CV states are generated by initial interaction of even state $|\Omega_+^{(l)}\rangle$ with the delocalized photon. The meaning of subscripts and superscripts in the amplitudes $x_{lp}^{(2m+)}$ and $y_{lp}^{(2m+)}$ is different. The second number $p$ of the lower index is responsible for the summation in Eqs. (15,16) and the first number $l$ coincides with the number of the initial CV state in Eq. (1). The superscript $(2m +)$ indicates that the amplitudes are the result of the interaction of even CV state $|\Omega_+^{(l)}\rangle$ with the delocalized photon provided that even number of photons $2m$ is detected in the auxiliary mode. We are going to call the state like $|\Psi_{2m}^{(l+)}\rangle$ even CV one as it exclusively consist of even Fock states. The state like $|\Phi_{2m}^{(l+)}\rangle$ is solely composed of odd Fock states, therefore, it can be called odd CV state. It is worth noting that the state (15) includes $l + 1$ superposition terms while the state (16) comprises $l + 2$ terms.

All other possible CV states can be written in the same form as in equations (15,16) with the corresponding notations that specify the initial conditions and the measurement outcomes. All possible combinations of the parities of the conditional CV states depending on the choice of the initial state and the measured outcome are shown in Table 1. The analytical form of the success probabilities $P_{2m}^{(l\pm)}$ and $P_{2m+1}^{(l\pm)}$ to generate the hybrid entangled states is presented in Appendixes B and C. Note that the common normalization factors $\mathfrak{N}_{2m}^{(l\pm)} = \left(|a_0|^2 + |a_1|^2 \left|B_{2m}^{(l\pm)}\right|^2\right)^{-1/2}$ and $\mathfrak{N}_{2m+1}^{(l\pm)} = \left(|a_0|^2 + |a_1|^2 \left|B_{2m+1}^{(l\pm)}\right|^2\right)^{-1/2}$ to a large extent determine the success probabilities of the generation of the hybrid entangled states in Eqs. (13,14).

| initial state | ($l+$) | | ($l-$) | |
| --- | --- | --- | --- | --- |
| | even | | odd | |
| measurement outcome | even | odd | even | odd |
| $|\Psi\rangle$ | even | odd | odd | even |
| $|\Phi\rangle$ | Odd | even | even | odd |

**Table 1.** The parity of the CV states $|\Psi\rangle$ and $|\Phi\rangle$ being part of the hybrid entangled states in Eqs. (13,14) in dependency on parity of input CV state in Eq. (1) and parity of the measurement outcome detected.

## IV. ENTANGLEMENT OF THE GENERATED STATES

The hybrid entangled states in Eqs. (13,14) exist in the four-dimensional Hilbert space. Indeed, regardless of the input CV state, the BS parameters and measurement outcomes, only a set of orthogonal states $\{|even\rangle_1|0\rangle_2, |odd\rangle_1|0\rangle_2, |even\rangle_1|1\rangle_2, |odd\rangle_1|1\rangle_2\}$ can be basis states of the four-dimensional Hilbert space, where by designations $|even\rangle, |odd\rangle$ it is meant the states that exclusively contain either even or odd Fock states. In particular, the following the Table 1 the states $|\Psi_{2m}^{(l+)}\rangle$, $|\Psi_{2m+1}^{(l-)}\rangle$, $|\Phi_{2m+1}^{(l+)}\rangle$ and $|\Phi_{2m}^{(l-)}\rangle$ can be recognized by $|even\rangle = \left\{|\Psi_{2m}^{(l+)}\rangle, |\Psi_{2m+1}^{(l-)}\rangle, |\Phi_{2m+1}^{(l+)}\rangle, |\Phi_{2m}^{(l-)}\rangle\right\}$, while the states $|\Psi_{2m+1}^{(l+)}\rangle$, $|\Psi_{2m}^{(l-)}\rangle$, $|\Phi_{2m}^{(l+)}\rangle$ and $|\Phi_{2m+1}^{(l-)}\rangle$ are the $|odd\rangle = \left\{|\Psi_{2m+1}^{(l+)}\rangle, |\Psi_{2m}^{(l-)}\rangle, |\Phi_{2m}^{(l+)}\rangle, |\Phi_{2m+1}^{(l-)}\rangle\right\}$ states. The CV states $|even\rangle$ and $|odd\rangle$ are orthogonal to each other $\langle odd|even\rangle = 0$ for each pair taken from the set of even and odd states.

Measure of the entanglement of the conditional entangled states in Eqs. (13,14) can be estimated by using partial transpose (PPT) criterion for separability [33,34]. The negativity $\mathcal{N}$ has all required properties for the entanglement measure. The negativity value ranges from $\mathcal{N}_s = 0$ (separable state) up to $\mathcal{N}_{max} = 1$ (maximally entangled state). One can calculate the negativities $\mathcal{N}_{2m}^{(l\pm)}$ and $\mathcal{N}_{2m+1}^{(l\pm)}$, respectively, in four-dimensional Hilbert space

$$\mathcal{N}_{2m}^{(l\pm)} = \frac{2|a_0||a_1|\left|B_{2m}^{(l\pm)}\right|}{|a_0|^2 + |a_1|^2\left|B_{2m}^{(l\pm)}\right|^2}, \qquad (17)$$

$$\mathcal{N}_{2m+1}^{(l\pm)} = \frac{2|a_0||a_1|\left|B_{2m+1}^{(l\pm)}\right|}{|a_0|^2 + |a_1|^2\left|B_{2m+1}^{(l\pm)}\right|^2}. \tag{18}$$

The hybrid entangled states in Eqs. (13,14) contain an additional parameter $B_{2m}^{(l\pm)}$ or $B_{2m+1}^{(l\pm)}$, respectively, the appearance of which is associated with the features of the interaction of an arbitrary input state $|\Omega_{\pm}^{(l)}\rangle$ with the delocalized photon. The analytical form of the parameter, that largely determines the measure of entanglement of the conditioned states, is presented in the Appendixes B and C. The parameter is entirely determined by input experimental conditions.

As can be seen from the expressions, negativity can take on zero value only if either $B_{2m}^{(l\pm)}$ or $B_{2m+1}^{(l\pm)}$ are equal to zero since it is initially assumed that the amplitudes $a_0, a_1$ of the delocalized photon take non-zero values. Thus, the conditional states in Eqs. (13,14) always have some degree of entanglement unless $B_{2m}^{(l\pm)} \neq 0$ and $B_{2m+1}^{(l\pm)} \neq 0$. The maximum degree of the negativity is observed provided that the condition either $|a_0| = |a_1|\left|B_{2m}^{(l\pm)}\right|$ or $|a_0| = |a_1|\left|B_{2m+1}^{(l\pm)}\right|$ is implemented. Maximally entangled states can be generated if the balanced delocalized photon $\left(|a_0| = |a_1| = 1/\sqrt{2}\right)$ is used in the case of $\left|B_{2m}^{(l\pm)}\right| = 1$ and $\left|B_{2m+1}^{(l\pm)}\right| = 1$. As can be seen from the analytical expressions for the parameters $B_{2m}^{(l\pm)}$ in Eqs. (B12,B19,C6, D8) and $B_{2m+1}^{(l\pm)}$ in Eqs. (B13,B20,C11), they are completely determined by the values of initial experimental parameters. This means that the parameters never take zero value $\left(B_{2m}^{(l\pm)} \neq 0, B_{2m+1}^{(l\pm)} \neq 0\right)$ which, in combination with the condition for the existence of the nonlocalized photon ($a_0 \neq 0, a_1 \neq 0$), makes it possible to say of deterministic generation of the entangled hybridity for all possible input states in Eq. (1) and used experimental parameters.

We show the plots of the negativities $\left(\mathcal{N}_0^{(0+)}, \mathcal{N}_1^{(0+)}, \mathcal{N}_0^{(1+)}, \mathcal{N}_1^{(1+)}\right)$ (left side plots) and corresponding them success probabilities $\left(P_0^{(0+)}, P_1^{(0+)}, P_0^{(1+)}, P_1^{(1+)}\right)$ (graphs on the right side) to generate the states for even initial states $|\Omega_+^{(0)}\rangle$ and $|\Omega_+^{(1)}\rangle$ in dependency on $\beta$ and $t$ in Figure 6. As can be seen from the plots, there is a fairly large range of values $(\beta, t)$ in which negativity can take on rather large values close to maximal $\mathcal{N}_{max} = 1$. Note that the success probabilities can also take rather large values in the given range of experimental parameters. In Figure 7, we also show the negativities $\left(\mathcal{N}_0^{(0-)}, \mathcal{N}_1^{(0-)}, \mathcal{N}_0^{(1-)}, \mathcal{N}_1^{(1-)}\right)$ (on the left side) and their corresponding success probabilities $\left(P_0^{(0-)}, P_1^{(0-)}, P_0^{(1-)}, P_1^{(1-)}\right)$ (on the right side) as functions of the parameters $\beta$ and $t$ for input odd CV states $|\Omega_-^{(0)}\rangle$ and $|\Omega_-^{(1)}\rangle$. They also have areas of parameters $(\beta, t)$ in which the negativity can take values close to $\mathcal{N}_{max} = 1$. The plots are constructed for case of the balanced delocalized photon $a_0 = a_1 = 1/\sqrt{2}$.

A more general case involves the use of the even/odd CV states in Eqs. (9,10). Due to the linearity of the BS operator, the output state can be written as

$$BS_{12}\left(|\Omega_{in}^{(\pm)}\rangle_1 |\varphi\rangle_{23}\right) = N_l^{(\pm)} \sum_{k=0}^{l} b_k^{(\pm)} BS_{12}\left(|\Omega_{\pm}^{(k)}\rangle_1 |\varphi\rangle_{23}\right). \tag{19}$$

Each term $BS_{12}\left(|\Omega_+^{(k)}\rangle_1 |\varphi\rangle_{23}\right)$ contributes to the generated entanglement. Finally, the contributions are summed up and the conditional state will have a rather complex form as shown in Appendix D. Nevertheless, the final conditional state can be represented by the state of the same form as in Eq. (13) in the case of even number of photons $n = 2m$ detected. In the case of an odd number $n = 2m + 1$ photons detected, the conditioned state can be represented in the same form as in the equation (14). The parity of the CV states of the generated entangled states follows from Table 1. The conditional states can also be described in a four-dimensional Hilbert space regardless of the parity of the measured photons in the auxiliary mode. Thus, one can also use expressions (17,18) for calculating the negativity of the conditioned states in the case of input superposition (19). In general, calculating negativity and success probability for an arbitrary number $l$ superposition terms in Eq. (19) is quite difficult and tedious. In a particular case of $l = 1$ $\left(|even\rangle \equiv |\Omega_{in}^{(+)}\rangle = N_1^{(+)}\left(|\Omega_+^{(0)}\rangle + |\Omega_+^{(1)}\rangle\right)\right.$, $\left.|odd\rangle \equiv |\Omega_{in}^{(-)}\rangle = |odd\rangle = N_1^{(-)}\left(|\Omega_-^{(0)}\rangle + |\Omega_-^{(1)}\rangle\right)\right)$ we calculated the negativities $\left(\mathcal{N}_0^{([0,1]+)}, \mathcal{N}_1^{([0,1]+)}, \mathcal{N}_0^{([0,1]-)}, \mathcal{N}_1^{([0,1]-)}\right)$ of the conditional states and corresponding them success probabilities $\left(P_0^{([0,1]+)}, P_1^{([0,1]+)}, P_0^{([0,1]-)}, P_1^{([0,1]-)}\right)$. Plots of the parameters are presented in Figure 8 in dependency on experimental parameters $\beta$ and $t$ in the case of $a_0 = a_1 = 1/\sqrt{2}$. It is interesting to note that a fairly smooth shape is observed for the negativities $\mathcal{N}_0^{([0,1]+)}$ and $\mathcal{N}_0^{([0,1]-)}$, while the shape of surfaces $\mathcal{N}_1^{([0,1]+)}$ and $\mathcal{N}_1^{([0,1]-)}$ have sharp drops.

Numerical simulation shows that the number of parameter values at which the maximum negativity $\mathcal{N}_{max}$ is observed is very large. Some values of the experimental parameters $(\beta, t)$ at which the maximum negativity $\mathcal{N}_{max}$ is observed are presented in Table 2 for the case of balanced delocalized photon $a_0 = a_1 = 1/\sqrt{2}$. Note that numerical calculations, which we do not present here, show that the maximum entanglement is also observed in the case of an unbalanced delocalized photon $a_0 \neq a_1$ in a large number of cases.

| Initial state | Clicks | Probability | $\beta$ | $t$ |
|---|---|---|---|---|
| $\left\vert\Omega_+^{(0)}\right\rangle$ | 0 | 0.939 | 0.5 | 0.25 |
| $\left\vert\Omega_+^{(0)}\right\rangle$ | 1 | 0.288 | 1.4 | 0.65 |
| $\left\vert\Omega_+^{(1)}\right\rangle$ | 0 | 0.491 | 0.5 | 0.73 |
| $\left\vert\Omega_+^{(1)}\right\rangle$ | 1 | 0.301 | 0.5 | 0.61 |
| $\left\vert\Omega_-^{(0)}\right\rangle$ | 0 | 0.544 | 0.5 | 0.79 |
| $\left\vert\Omega_-^{(0)}\right\rangle$ | 1 | 0.843 | 0.5 | 0.25 |
| $\left\vert\Omega_-^{(1)}\right\rangle$ | 0 | 0.523 | 0.5 | 0.8 |
| $\left\vert\Omega_-^{(1)}\right\rangle$ | 1 | 0.278 | 2.1 | 0.96 |
| $|even\rangle$ | 0 | 0.938 | 0.92 | 0.25 |
| $|even\rangle$ | 1 | 0.291 | 1.9 | 0.62 |
| $|odd\rangle$ | 0 | 0.509 | 1.34 | 0.8 |
| $|odd\rangle$ | 1 | 0.31 | 0.5 | 0.68 |

**Table 2.** Values of the experimental parameters $(\beta, t)$, at which the maximum negativity $\mathcal{N}_{max} = 1$ of generated state in Figure 5 is observed. Corresponding success probabilities are also presented. Here, the following notations $|even\rangle = N_1^{(+)}\left(|\Omega_+^{(0)}\rangle + |\Omega_+^{(1)}\rangle\right)$ and $|odd\rangle = N_1^{(-)}\left(|\Omega_-^{(0)}\rangle + |\Omega_-^{(1)}\rangle\right)$ are used.

## V. TRUNCATED VERSIONS OF THE INPUT CV STATES

It is usually quite difficult to implement the SDlPSs in practice since most likely the generation of the states will require extreme values of Kerr (cubic) nonlinearity, which is currently unattainable. So, in the case of the SCSs generation, the cubic nonlinearity of existing optical media is insufficient, taking into account the fact that the incipient SCSs is subjected to the effect of decoherence along its propagation, which damps its coherence. As a rule, instead of the CV states, their truncated versions are used, which include a limited number of terms in the superposition. In general case, the finite superpositions can be represented as

$$\left|\Omega_{in}^{(+l)}\right\rangle = N_+^{(l)} \sum_{m=0}^{l} d_{2m}^{(+)} |2m\rangle, \qquad (20)$$

in the case of original even state and

$$\left|\Omega_{in}^{(-l)}\right\rangle = N_-^{(l)} \sum_{m=0}^{l} d_{2m+1}^{(+)} |2m+1\rangle, \qquad (21)$$

in the case of input odd state, where $N_\pm^{(l)}$ are the corresponding normalization factors. If the amplitudes take on values $d_{2m}^{(+)} = c_{l\,2m}(\beta)$, then we deal with truncated version of the even SDlPSs in Eq. (5) with some value of $l$. If $d_{2m+1}^{(-)} = c_{l\,2m+1}(\beta)$, then we use the truncated version of odd SDlPSs in Eq. (6).

The truncated versions of the even and odd CV states can also be used in the optical scheme in Figure 5. An entangled state is also generated in the case of registration of any measurement outcome in the second auxiliary mode excluding $2m+1$ measurement outcome in the case of the input state (20) and $2m+2$ measurement outcome in the case of the input state (21). The conditional entangled state can also be described in four-dimensional Hilbert space as in the case discussed above with input CV states. In other words, the generated entangled states will have the same form as in the equations (13,14). The only difference is that the CV states forming the entanglement are replaced by the finite superpositions. The negativity of the entanglement is calculated by the equations (17,18). Numerical calculations show that the resulting maximum entanglement $\mathcal{N}_{max} = 1$ is also observed in a large number of the experimental parameters used.

## VI. CONCLUSION

We considered a family of the CV states being the superposition of the DNSs with equal modulus but different in sign displacement amplitudes. The family of the CV states is a generalization of the well-known SCSs being optical analogue of Schrödinger cat states [27]. As in the case of SCSs, the SDlPSs (1) are divided into even and odd depending on the parity of the Fock states forming a superposition. We constucted the Wigner functions some of the SDlPSs in Fig. 1 and showed that they have inherent nonclassical properties like manifestation of interference and regions on phase plane, where the Wigner functions take on negative values. It should be noted that separated peaks are observed in one of the quadrature distributions of the studied states (Fig. 2), as well as in the case of the SCSs. The nonclassical properties of the SDlPSs can also manifest in the observation of quadrature squeezing and also in the detection of the Fano factor taking values less than 1.

The SDlPSs can be directly used to generate the conditional entanglement in the equations (13,14). The explanation of this result can be traced to the example of even SDlPSs and even measurement outcome $2m$. If even number of photons comes from even SDlPSs, then heralded state can only comprise even Fock states as even number of photons is detected at auxiliary mode. In order case, if even Fock states of the even SDlPSs are mixed with single photon, the resulting state can only involve odd Fock states in the case of registration of even number of photons in auxiliary mode. Due to indistinguishability of the events, conditional hybrid entangled state in Eq. (13) is obtained. The same explanation applies to the three remaining cases characterized by the parity of the input and the measurement outcomes. The CV components of the heralded entangled state acquire the parity as noted in Table 1. This explanation is also applicable to input superpositions consisted of SDlPSs and the truncated versions of the states. The generated states have some degree of entanglement characterized by the negativity. Negativity is largely determined by the parameter $B_{2m}^{(l\pm)}, B_{2m+1}^{(l\pm)}$ occurring due interaction of multiphoton states at the beam splitter. This parameter always takes nonzero values, which indicates the possibility of deterministic generation of hybrid entanglement. At certain values of the experimental parameters, the negativity of the generated entanglement takes on the maximum possible value. The set of such parameters that ensure the generation of maximally entangled state is large. In addition, the experimental parameters can be selected in such a way to provide a sufficiently high success probability of the hybrid entanglement generation. The method of entanglement generation can be used to generate entangled multipartite states such as graph states or cluster states and in perspective large-scale quantum networks. This can be done through sequential spreading the entanglement between parts of the incipient multipartite state in the same manner as was considered in the case of two states.

## APPENDIX A. NOTES ABOUT DNSS

Consider the DNS in the number states basis [35,36]

$$|n,\alpha\rangle \equiv D(\alpha)|n\rangle = F(\alpha)\sum_{m=0}^{\infty} c_{nm}(\alpha)|m\rangle, \quad (A1)$$

where the unitary displacement operator is $D(\beta) = exp(\beta a^+ - \beta^* a)$ with amplitude $\beta$ and $a\ (a^+)$ are bosonic annihilation (creation) operators. The normalization factor $F(\alpha)$ is introduced above. The amplitudes $c_{nm}(\alpha)$ are calculated as

$$c_{nm}(\alpha) = exp(|\alpha|^2/2)\langle m|n,\alpha\rangle, \quad (A2)$$

that provides normalization condition $exp(-|\alpha|^2)\sum_{m=0}^{\infty} c_{lm}^*(\alpha)c_{nm}(\alpha) = \delta_{ln}$ for any numbers $l$ and $n$, where $\delta_{ln} = 1$ if $l = n$ and $\delta_{ln} = 0$ if $l \neq n$. It can be shown in [35], the following relation holds

$$c_{nm}(-\alpha) = (-1)^{m-n}c_{nm}(\alpha). \quad (A3)$$

## APPENDIX B. INTERACTION OF SCSS WITH DELOCALIZED PHOTON

Consider the interaction of the even SCSs $|\Omega_+^{(0)}\rangle$ in Eq. (3) with the second mode of the delocalized photon in Eq. (11) on the beam splitter (12). Due to the linearity of the beam splitter operator, we have

$$BS_{12}\left(\left|\Omega_+^{(0)}\right\rangle_1 |\varphi\rangle_{23}\right) = N_+^{(0)}(\beta)\left(BS_{12}(|-\beta\rangle_1|\varphi\rangle_{23}) + BS_{12}(|\beta\rangle_1|\varphi\rangle_{23})\right). \quad (B1)$$

Let us consider the output state in the case of mixing of the coherent state $|-\beta\rangle_1$ with the delocalized photon (12) on the BS

$$BS_{12}(|-\beta\rangle_1|\varphi\rangle_{23}) = BS_{12}D(-\beta)(|0\rangle_1|\varphi\rangle_{23}) = BS_{12}D_1(-\beta)BS_{12}^+BS_{12}(|0\rangle_1|\varphi\rangle_{23}) =$$
$$D_1(-\beta t)D_2(\beta r)(a_0|00\rangle_{12}|1\rangle_3 + a_1(t|01\rangle_{12} + r|10\rangle_{12})|0\rangle_3) =$$
$$(a_0|0,-\beta t\rangle_1|0,\beta r\rangle_2|1\rangle_3 + a_1(t|0,-\beta t\rangle_1|1,\beta r\rangle_2 + r|1,-\beta t\rangle_1|0,\beta r\rangle_2))|0\rangle_3, \quad (B2)$$

where we embraced by unitarity of the beam splitter operator $BS_{12}BS_{12}^+ = BS_{12}^+BS_{12} = I$ with $I$ being identical operator. The same transformations apply to the state $BS_{12}(|\beta\rangle_1|\varphi\rangle_{23})$

$$BS_{12}(|\beta\rangle_1|\varphi\rangle_{23}) = BS_{12}D(\beta)(|0\rangle_1|\varphi\rangle_{23}) = BS_{12}D_1(\beta)BS_{12}^+BS_{12}(|0\rangle_1|\varphi\rangle_{23}) =$$
$$D_1(\beta t)D_1(-\beta r)(a_0|00\rangle_{12}|1\rangle_3 + a_1(t|01\rangle_{12} + r|10\rangle_{12})|0\rangle_3) =$$

$$(a_0|0,\beta t\rangle_1|0,-\beta r\rangle_2|1\rangle_3 + a_1(t|0,\beta t\rangle_1|1,-\beta r\rangle_2 + r|1,\beta t\rangle_1|0,-\beta r\rangle_2))|0\rangle_3. \qquad (B3)$$

Using the equations (B1)-(B3), we can write the final expression for the output state

$$BS_{12}\left(|\Omega_+^{(0)}\rangle_1 |\varphi\rangle_{23}\right) = N_+^{(0)}(\beta)\big(a_0(|0,-\beta t\rangle_1|0,\beta r\rangle_2 + |0,\beta t\rangle_1|0,-\beta r\rangle_2)|1\rangle_3 +$$
$$a_1\big(t(|0,-\beta t\rangle_1|1,\beta r\rangle_2 + |0,\beta t\rangle_1|1,-\beta r\rangle_2) + r(|1,-\beta t\rangle_1|0,\beta r\rangle_2 +$$
$$|1,\beta t\rangle_1|0,-\beta r\rangle_2)\big)|0\rangle_3\big). \qquad (B4)$$

Now we can use the decomposition of the displaced states in the Fock basis (A1) taking into account the properties of the matrix elements when changing the sign of the displacement amplitude $\alpha$ to the opposite $\alpha \to -\alpha$ given by Eq. (A3)

$$BS_{12}\left(|\Omega_+^{(0)}\rangle_1 |\varphi\rangle_{23}\right) = N_+^{(0)}(\beta)F(\beta r)\sum_{n=0}^{\infty}\big(a_0 c_{0n}(\beta r)(|0,-\beta t\rangle_1 + (-1)^n|0,\beta t\rangle_1)|1\rangle_3 +$$
$$a_1\big(tc_{1n}(\beta r)(|0,-\beta t\rangle_1 + (-1)^{n-1}|0,\beta t\rangle_1) +$$
$$rc_{0n}(\beta r)(|1,-\beta t\rangle_1 + (-1)^n|1,\beta t\rangle_1)\big)|0\rangle_3\big)|n\rangle_2, \qquad (B5)$$

Measurement outcomes in the second mode can be divided into two types depending on the parity of the measured photons: either even $n = 2m$ or odd $n = 2m + 1$ photons detected. So, if even number of photons $2m$ is registered in second mode, then the hybrid entangled state in Eq. (13) is generated with the CV states in Eqs. (15)-(16) whose amplitudes are the following

$$x_{00}^{(2m+)} = 1, \qquad (B6)$$
$$y_{00}^{(2m+)} = 1, \qquad (B7)$$
$$y_{01}^{(2m+)} = \frac{rc_{02m}(\beta r)N_-^{(0)}(\beta t)}{tc_{12m}(\beta r)N_+^{(1)}(\beta t)}. \qquad (B8)$$

In the case of registration of odd photons $2m + 1$ in auxiliary second mode, the hybrid entangled state in Eq. (14) is produced whose CV states have the following amplitudes

$$x_{00}^{(2m+1+)} = 1, \qquad (B9)$$
$$y_{00}^{(2m+1+)} = 1, \qquad (B10)$$
$$y_{01}^{(2m+)} = \frac{rc_{02m+1}(\beta r)N_+^{(0)}(\beta t)}{tc_{12m+1}(\beta r)N_-^{(1)}(\beta t)}. \qquad (B11)$$

The parameters $B_{2m}^{(0+)}$ and $B_{2m+1}^{(0+)}$ which to a large extent defines negativity are given by

$$B_{2m}^{(0+)} = \frac{tc_{12m}(\beta r)N_-^{(0)}(\beta t)}{c_{02m}(\beta r)N_-^{(0)}(\beta t)K_{2m}^{(0+)}}, \qquad (B12)$$

$$B_{2m+1}^{(0+)} = \frac{tc_{12m+1}(\beta r)N_-^{(0)}(\beta t)}{c_{02m+1}(\beta r)N_+^{(0)}(\beta t)K_{2m+1}^{(0+)}}, \qquad (B13)$$

where all values used are given above. The success probabilities to generate the conditional hybrid entangled states in Eq. (13)-(14) are the following

$$P_{2m}^{(0+)} = \frac{F^2(\beta r)|c_{02m}(\beta r)|^2 N_+^{(0)2}(\beta)}{N_+^{(0)2}(\beta t)\mathfrak{N}_{2m}^{(0+)2}}, \qquad (B14)$$

$$P_{2m+1}^{(0+)} = \frac{F^2(\beta r)|c_{02m+1}(\beta r)|^2 N_+^{(0)2}(\beta)}{N_-^{(0)2}(\beta t)\mathfrak{N}_{2m+1}^{(0+)2}}, \qquad (B15)$$

The same consideration applies to the state $|\Omega_-^{(0)}\rangle$ in Eq. (3) and resulting states become either in Eq. (13) or Eq. (14) in dependency on the measurement outcomes. Summarizing the above analysis, it is possible to write

$$x_{00}^{(2m-)} = x_{00}^{(2m+1-)} = y_{00}^{(2m-)} = y_{00}^{(2m-)} = 1, \qquad (B16)$$

$$y_{01}^{(2m-)} = \frac{rc_{02m}(\beta r)N_+^{(0)}(\beta t)}{tc_{12m}(\beta r)N_-^{(1)}(\beta t)}, \qquad (B17)$$

$$y_{01}^{(2m+1+)} = \frac{rc_{02m+1}(\beta r)N_-^{(0)}(\beta t)}{tc_{12m+1}(\beta r)N_+^{(1)}(\beta t)}. \qquad (B18)$$

The parameters $B_{2m}^{(0-)}$ and $B_{2m+1}^{(0-)}$ become

$$B_{2m}^{(0-)} = \frac{tc_{12m}(\beta r)N_-^{(0)}(\beta t)}{c_{02m}(\beta r)N_+^{(0)}(\beta t)K_{2m}^{(0-)}}, \tag{B19}$$

$$B_{2m+1}^{(0-)} = \frac{tc_{12m+1}(\beta r)N_+^{(0)}(\beta t)}{c_{02m+1}(\beta r)N_-^{(0)}(\beta t)K_{2m+1}^{(0-)}}. \tag{B20}$$

while the success probabilities are the following

$$P_{2m}^{(0-)} = \frac{F^2(\beta r)|c_{02m}(\beta r)|^2 N_-^{(0)2}(\beta)}{N_-^{(0)2}(\beta t)\mathfrak{N}_{2m}^{(0-)2}}, \tag{B21}$$

$$P_{2m+1}^{(0-)} = \frac{F^2(\beta r)|c_{02m+1}(\beta r)|^2 N_-^{(0)2}(\beta)}{N_+^{(0)2}(\beta t)\mathfrak{N}_{2m+1}^{(0-)2}}. \tag{B22}$$

By direct summation, it can be shown that the probabilities are normalized, i.e. $\sum_{m=0}^{\infty}\left(P_{2m}^{(0\pm)} + P_{2m+1}^{(0\pm)}\right) = 1$.

## APPENDIX C. INTERACTION OF EVEN/ODD CV STATES DELOCALIZED PHOTON (GENERAL CASE)

Now, we are going to derive the expressions (13)-(14) with corresponding CV whose parity is reflected in the Table 1. To do it consider the result of mixing of $l-$photons with vacuum and single photon, respectively, on the BS in Eq. (12)

$$BS_{12}(|l\rangle_1|0\rangle_2) = \sum_{k=0}^{l}(-1)^k t^{l-k}r^k \sqrt{\frac{l!}{k!(l-k)!}}|l-k\rangle_1|k\rangle_2, \tag{C1}$$

$$BS_{12}(|l\rangle_1|1\rangle_2) = \sqrt{l+1}t^l r|l+1\rangle_1|0\rangle_2 +$$
$$\sum_{k=0}^{l}(-1)^k \frac{t^{l-k-1}r^k}{k!}\sqrt{\frac{(k+1)!l!}{(l-k)!}}\left(t^2 - \frac{l-k}{k+1}r^2\right)|l-k\rangle_1|k+1\rangle_2. \tag{C2}$$

The states are the basis for the derivation of the conditional states in Eqs. (13)-(14).

Consider the even state $|\Omega_+^{(l)}\rangle$ in Eq. (1) with arbitrary value of $l$. Using the same calculation technique that was already used in the derivation of both $|\Delta_{2m}^{(0+)}\rangle_{13}$ and $|\Delta_{2m+1}^{(0+)}\rangle_{13}$, it is possible to show the hybrid entangled state in Eq. (13) is conditionally generated provided that $n = 2m$ photons are fixed in second auxiliary mode with corresponding CV states even $|\Psi_{2m}^{(l+)}\rangle$ and odd $|\Phi_{2m}^{(l+)}\rangle$ given by Eqs. (15)-(16) whose amplitudes are the following

$$x_{lp}^{(2m+)} = (-1)^p \left(\frac{t}{r}\right)^p \sqrt{\frac{l!}{p!(l-p)!}} \frac{c_{l-p2m}(\beta r)N_+^{(0)}(\beta t)}{c_{l2m}(\beta r)N_+^{(p)}(\beta t)}, \tag{C3}$$

$$y_{lp}^{(2m+)} = (-1)^p \frac{t^{p-2}\sqrt{l!(l-p+1)!}c_{l-p+12m}(\beta r)N_-^{(0)}(\beta t)}{r^p(l-p)!\sqrt{(l+1)p!}c_{l+12m}(\beta r)N_-^{(p)}(\beta t)}\left(t^2 - \frac{p}{l-p+1}r^2\right), \tag{C4}$$

$$y_{ll+1}^{(2m+)} = (-1)^l \frac{t^{l-1}c_{02m}(\beta r)N_-^{(0)}(\beta t)}{r^{l-1}c_{l+12m}(\beta r)N_-^{(l+1)}(\beta t)}. \tag{C5}$$

Rather tedious calculations allow for one to find an analytical expression for the entanglement parameter $B_{2m}^{(l+)}$ and the success probability $P_{2m}^{(l+)}$ to generate the conditional state (13)

$$B_{2m}^{(l+)} = \frac{t\sqrt{(l+1)}c_{l+12m}(\beta r)N_+^{(0)}(\beta t)L_{2m}^{(l+)}(\beta t)}{c_{l2m}(\beta r)N_-^{(0)}(\beta t)K_{2m}^{(l+)}(\beta t)}, \tag{C6}$$

$$P_{2m}^{(l+)} = \frac{F^2(\beta r)|r|^{2l}|c_{l2m}(\beta r)|^2 N_+^{(l)2}(\beta)}{N_+^{(0)2}(\beta t)L_{2m}^{(l+)2}(\beta t)\mathfrak{N}_{2m}^{(l+)2}}, \tag{C7}$$

where $L_{2m}^{(l+)}$, $K_{2m}^{(l+)}$ and $\mathfrak{N}_{2m}^{(l+)}$ are the normalization factors of the states $|\Psi_{2m}^{(l+)}\rangle$, $|\Phi_{2m}^{(l+)}\rangle$ and $|\Delta_{2m}^{(l+)}\rangle$, respectively.

By analogy, one can obtain analytical expressions for the parameters for the entangled state in the case of odd measurement outcome $2m+1$. It is possible to show the conditional state is given in equation (14) with odd $|\Psi_{2m+1}^{(l+)}\rangle$ and even $|\Phi_{2m+1}^{(l+)}\rangle$ states whose amplitudes are the following

$$x_{lp}^{(2m+1+)} = (-1)^p \left(\frac{t}{r}\right)^p \sqrt{\frac{l!}{p!(l-p)!}} \frac{c_{l-p2m+1}(\beta r) N_-^{(0)}(\beta t)}{c_{l2m+1}(\beta r) N_-^{(p)}(\beta t)}, \tag{C8}$$

$$y_{lp}^{(2m+1+)} = (-1)^p \frac{t^{p-2}\sqrt{l!(l-p+1)!} c_{l-p+12m+1}(\beta r) N_+^{(0)}(\beta t)}{r^p (l-p)! \sqrt{(l+1)p!} c_{l+12m+1}(\beta r) N_+^{(p)}(\beta t)} \left(t^2 - \frac{p}{l-p+1} r^2\right), \tag{C9}$$

$$y_{ll+1}^{(2m+1+)} = (-1)^l \frac{t^{l-1} c_{02m+1}(\beta r) N_+^{(0)}(\beta t)}{r^{l-1} c_{l+12m+1}(\beta r) N_+^{(l+1)}(\beta t)}. \tag{C10}$$

The entanglement parameter $B_{2m+1}^{(l+)}$ and the success probability $P_{2m+1}^{(l+)}$ to generate the conditional state (14) are given by

$$B_{2m+1}^{(l+)} = \frac{t\sqrt{(l+1)} c_{l+12m+1}(\beta r) N_-^{(0)}(\beta t) L_{2m+1}^{(l+)}(\beta t)}{c_{l2m+1}(\beta r) N_+^{(0)}(\beta t) K_{2m+1}^{(l+)}(\beta t)}, \tag{C11}$$

$$P_{2m+1}^{(l+)} = \frac{F^2(\beta r) r^{2l} |c_{l2m+1}(\beta r)|^2 N_+^{(l)2}(\beta)}{N_-^{(0)2}(\beta t) L_{2m+1}^{(l+)2}(\beta t) \mathfrak{N}_{2m+1}^{(l+)2}}. \tag{C12}$$

If we consider the case of the initial CV state $|\Omega_-^{(l)}\rangle$ in Eq. (1), then the same calculation technique takes place. The difference will be only in some factors. Consider the difference on example of amplitudes in Eqs. (C3)-(C4). So, we must use the factor $N_-^{(0)}(\beta t)/N_-^{(p)}(\beta t)$ instead of $N_+^{(0)}(\beta t)/N_+^{(p)}(\beta t)$ in Eq. (C3) for the amplitudes $x_{lp}^{(2m-)}$. To obtain analytic expressions for amplitudes $y_{lp}^{(2m-)}$ and $y_{ll+1}^{(2m-)}$ from Eqs. (C4)-(C5), we must use the substitution $N_-^{(0)}(\beta t)/N_-^{(p)}(\beta t) \to N_+^{(0)}(\beta t)/N_+^{(p)}(\beta t)$ in Eq. (C4) and $N_-^{(0)}(\beta t)/N_-^{(l+1)}(\beta t) \to N_+^{(0)}(\beta t)/N_+^{(l+1)}(\beta t)$ in Eq. (C5). The corresponding changes should be made in the equations (C6)-(C7) in order to obtain analytical expressions for $B_{2m}^{(l-)}$ and $P_{2m}^{(l-)}$.

## APPENDIX D. NOTES ABOUT INTERACTION OF EVEN/ODD STATES (19) WITH DELOCALIZED PHOTON

To obtain analytical expressions for the amplitudes, it is worth making use of again the technique developed above. Consider it on example of input even CV state $|\Omega_{in}^{(+)}\rangle$ in Eq. (19) in the case of registration of even number $n=2m$ photons in second auxiliary mode. Calculations give the following amplitudes

$$x_p^{(2m)} = (-1)^p \frac{t^p}{\sqrt{p!}} \frac{f_{kp}^{(2m)} N_+^{(0)}(\beta t)}{f_{k0}^{(2m)} N_+^{(p)}(\beta t)}, \tag{D1}$$

for the even CV state $|\Psi_{2m}^{(+)}\rangle = L_{2m}^{(+)} \sum_{p=0}^{k} x_p^{(2m)} |\Omega_+^{(p)}\rangle$ with $L_{2m}^{(+)}$ being the normalization factor, where new parameters are introduced

$$f_{kp}^{(2m)} = \sum_{j=p}^{k} (-1)^j \frac{b_j^{(+)} N_+^{(j)}(\beta) r^{j-p} c_{j-p2m}(\beta r) \sqrt{j!}}{\sqrt{(j-p)!}}, \tag{D2}$$

$$f_{k0}^{(2m)} = \sum_{j=0}^{k} (-1)^j b_j^{(+)} N_+^{(j)}(\beta) r^j c_{j2m}(\beta r). \tag{D3}$$

The odd CV state is represented as
$$|\Phi_{2m}^{(-)}\rangle = K_{2m}^{(+)}\left(\sum_{p=0}^{k} x_p^{(2m)} |\Omega_+^{(p)}\rangle + \left(rtc_{02m}(\beta r)N_-^{(0)}(\beta t)/g_{s0}^{(2m)}\right)\sum_{l=0}^{k} g_l^{(2m)}\Omega_-^{(l+1)}\rangle\right),$$
where $K_{2m}^{(+)}$ is the normalization factor, with amplitudes

$$y_p^{(2m)} = (-1)^p \frac{t^p}{\sqrt{p!}} \frac{g_{kp}^{(2m)} N_-^{(0)}(\beta t)}{g_{k0}^{(2m)} N_-^{(p)}(\beta t)}, \tag{D4}$$

$$g_l^{(2m)} = b_l^{(+)} N_+^{(l)}(\beta) t^l \sqrt{l+1} N_-^{(l+1)-1}(\beta t), \tag{D5}$$

with the following parameters

$$g_{kp}^{(2m)} = \sum_{j=p}^{k}(-1)^j \frac{b_j^{(+)} N_+^{(j)}(\beta) r^{j-p} c_{j-p+12m}(\beta r)\sqrt{j!(j-p+1)!}}{(j-p)!}\left(t^2 - \frac{p}{j-p+1}r^2\right), \tag{D6}$$

$$g_{k0}^{(2m)} = t^2 \sum_{j=0}^{k}(-1)^j b_j^{(+)} N_+^{(j)}(\beta) r^j c_{j+12m}(\beta r)\sqrt{j+1}. \tag{D7}$$

The parameter $B_{2m}^{(+)}$ largely determining the entanglement of the generated state becomes

$$B_{2m}^{(+)} = \frac{g_{k0}^{(2m)} N_+^{(0)}(\beta t) L_{2m}^{(+)}}{t f_{k0}^{(2m)} N_-^{(0)}(\beta t) K_{2m}^{(+)}}. \tag{D8}$$

The success probability to conditionally produce the hybrid entangled states is the following

$$P_{2m}^{(+)} = \frac{F^2(\beta r)|r|^{2l}\left|f_{s0}^{(2m)}\right|^2 N_+^{(k)2}}{N_+^{(0)2}(\beta t) L_{2m}^{(l+)2}(\beta t) \mathfrak{R}_{2m}^{(l+)2}}, \tag{D9}$$

where the notations previously introduced are used.

It can be shown by direct calculations that the above expressions are transformed into those already introduced in the Appendix C in the case of if all amplitudes of the input state $|\Omega_{in}^{(+)}\rangle$ in Eq. (19) take zero values $b_j^{(+)} = 0$ with the exception of one $b_l^{(+)} = 1$. The results can be extended to the case of recording an odd number of measurement outcomes $n = 2m + 1$. In the same way, the conditional hybrid entangled states can be analyzed in the case of using the input state $|\Omega_{in}^{(-)}\rangle$ in Eq. (19).

**References**

[1] D. Kurzyk, Introduction to quantum Entanglement, Theoretical and Applied Informatics **24**, 135 (2012).
[2] E. Andersson and P. Öhberg (editors), Quantum Information and Coherence (Springer International Publishing, Switzerland, 2014).
[3] M. Horodecki, Entanglement measures, Quantum Inform. and Computation **1**, 3 (2001).
[4] A. Einstein, B. Podolsky and N. Rosen, Can quantum-mechanical description of physical reality be considered complete?, Phys. Rev. **47**, 777 (1935).
[5] J. Bell, On the Einstein Podolsky Rosen paradox, Physics **1**, 195 (1964).
[6] S. Freedman and J. Clauser, Experimental test of local hidden-variable theories, Phys. Rev. Lett. **28**, 938 (1972).
[7] A. Aspect, P. Grangier, and G. Roger, Experimental tests of realistic local theories via Bell's theorem, Phys. Rev. Lett. **47**, 460 (1981).
[8] A. Aspect, P. Grangier, and G. Roger, Experimental realization of Einstein-Podolsky-Rosen-Bohm *gedankenexperiment*: a new violation of Bell's inequalities, Phys. Rev. Lett. **49**, 91 (1982).
[9] G. Weihs, T. Jennewein, C. Simon, H. Weinfurter and A. Zeilinger, Violation of Bell's inequality under strict Einstein locality conditions, Phys. Rev. Lett **81**, 5039 (1998).
[10] B. Hensen, and et al., Loophole-free Bell inequality using electron spins separated by 1.3 kilometres, Nature **526**, 682 (2015).
[11] D. Rauch and et al., Cosmic Bell test using random measurement settings from high-redshift quasars, Phys. Rev. Lett. **121**, 080403 (2018).

[12] C. H. Bennett, and et al., Teleporting an unknown state via dual classical and Einstein-Podolsky-Rosen channels, **70**, 1895 (1993).
[13] D. Bouwmeester, and et al., Experimental quantum teleportation, Nature, **390**, 575 (1997).
[14] N. Lee, and et al., Teleporation of nonclassical wave packets of light, Science **332**, 330 (2011).
[15] L. Vaidman, Teleportation of quantum states, Phys. Rev. A, **49**, 1473 (1994).
[16] S. L. Braunstein, and H. J. Kimble, Teleportation of continuous quantum variables, Phys. Rev. Lett. **80**, 869 (1998).
[17] S. A. Podoshvedov, Quantum teleportation protocol with an assistant who prepares amplitude modulated unknown qubit, JOSA B **35**, 861-877 (2018).
[18] S. A. Podoshvedov, Efficient quantum teleportation of unknown qubit based on DV-CV interaction mechanism, Entropy **21**, 150 (2019).
[19] O. S. Magaña-Loaiza, and et al., Multiphoton quantum-state engineering using conditional measurements, npj Quantum Information 5:80 (2019).
[20] E. V. Mikheev, A. S. Pugin, D. A. Kuts, S. A. Podoshvedov, and N. B. An, Efficient production of large-size optical Schrödinger cat states, Scientific Reports **9**, 14301 (2019).
[21] D. Gottesman, and I. L. Chuang, Demonstrating the viability of universal quantum computation using teleportation and single-qubit operations, Nature **402**, 390 (1999).
[22] R. Rausendorf, and H. J Briegel, A one-way quantum computer, Phys. Rev. Lett. **86**, 5188 (2001).
[23] M. A. Nielsen, and I. L. Chuang, Quantum Computation and Quantum Information (University Press, Cambridge, 2000).
[24] F. Arute, and et al., Quantum supremacy using a programmable superconducting processor, Nature **574**, 505 (2019).
[25] D. Burnham and D. Weinberg, Observation of simultaneity in parametric production of optical photon pairs, Phys. Rev. Lett **25**, 84 (1970).
[26] P. Kwiat, K. Mattle, H. Weinfurter, A. Zeilinger, A. Sergienko, and Y. Shih, New high-intensity source of polarization-entangled photon pairs, Phys. Rev. Lett. **75**, 4337 (1995).
[27] E. Schrödinger, Die gegenwärtige situation in der quantenmechanik, Naturwissenschaften **23**, 807 (1935).
[28] K. Huang, H. Le Jeannic, O. Morin, T. Darras, G. Guccione, A. Cavailles, and J. Laurat, Engineering optical hybrid entanglement between discrete- and continuous-variable states, New J. Phys. **21**, 083033 (2019).
[29] S. A. Podoshvedov and B. A. Nguyen, Designs of interactions between discrete- and continuous-variable states for generation of hybrid entanglement, Quantum Inf. Process. **18**, 68 (2019).
[30] M. S. Podoshvedov and S. A. Podoshvedov, Universal DV-CV interaction mechanism for deterministic generation of entangled hybridity, JOSA B, **37**, 963 (2020).
[31] A. E. Ulanov, D. Sychev, A. A. Pushkina, I. A. Fedorov and A. I. Lvovsky, Quantum teleportation between discrete and continuous encodings of an optical qubit, Phys. Rev. Lett. **118**, 160501 (2017).
[32] D. V. Sychev, A. E. Ulanov, E. S. Tiunov, A. A. Pushkina, A. Kuzhamuratov, V. Novikov and A. I. Lvovsky, Entanglement and teleportation between polarization and wave-like encodings of an optical qubit", Nat. Comm. **9**, 3672 (2018).
[33] A. Peres, Separability criterion for density matrices, Phys. Rev. Lett. **77**, 1413 (1996).
[34] G. Vidal, and R. F. Werner, Computable measure of entanglement, Phys. Rev. A **65**, 032314 (2002).
[35] S. A. Podoshvedov, Generation of displaced squeezed superpositions of coherent states, J. Exp. Theor. Phys. **114**, 451 (2012).

[36] S. A. Podoshvedov, Building of one-way Hadamard gate for squeezed coherent states. Phys. Rev. A **87**, 012307 (2013).
[37] D. N. Klyshko, The nonclassical light, Uspekhi Fizicheskikh Nauk, **39**, 573 (1996).

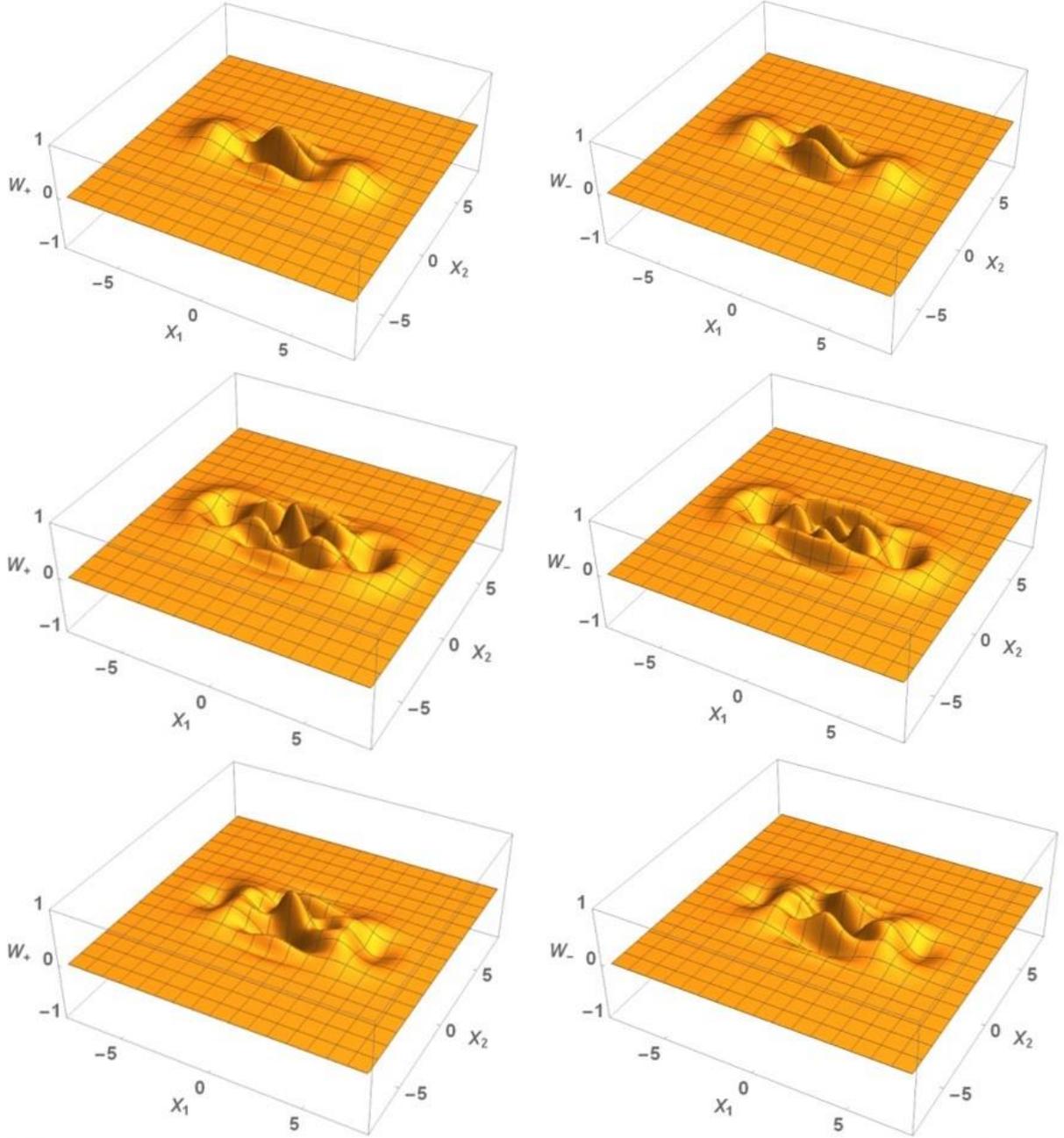

**FIG. 1.** Wigner functions $W_\pm$ for three states: even/odd SCSs $|\Omega_\pm^{(0)}\rangle$ (top two plots); even/odd SDSPSs $|\Omega_\pm^{(1)}\rangle$ (middle two plots) and CV superpostions: even $|even\rangle = N_1^{(+)}\left(|\Omega_+^{(0)}\rangle + |\Omega_+^{(1)}\rangle\right)$ and odd $|odd\rangle = N_1^{(-)}\left(|\Omega_-^{(0)}\rangle + |\Omega_-^{(1)}\rangle\right)$ (last two plots) for $\beta = 2$.

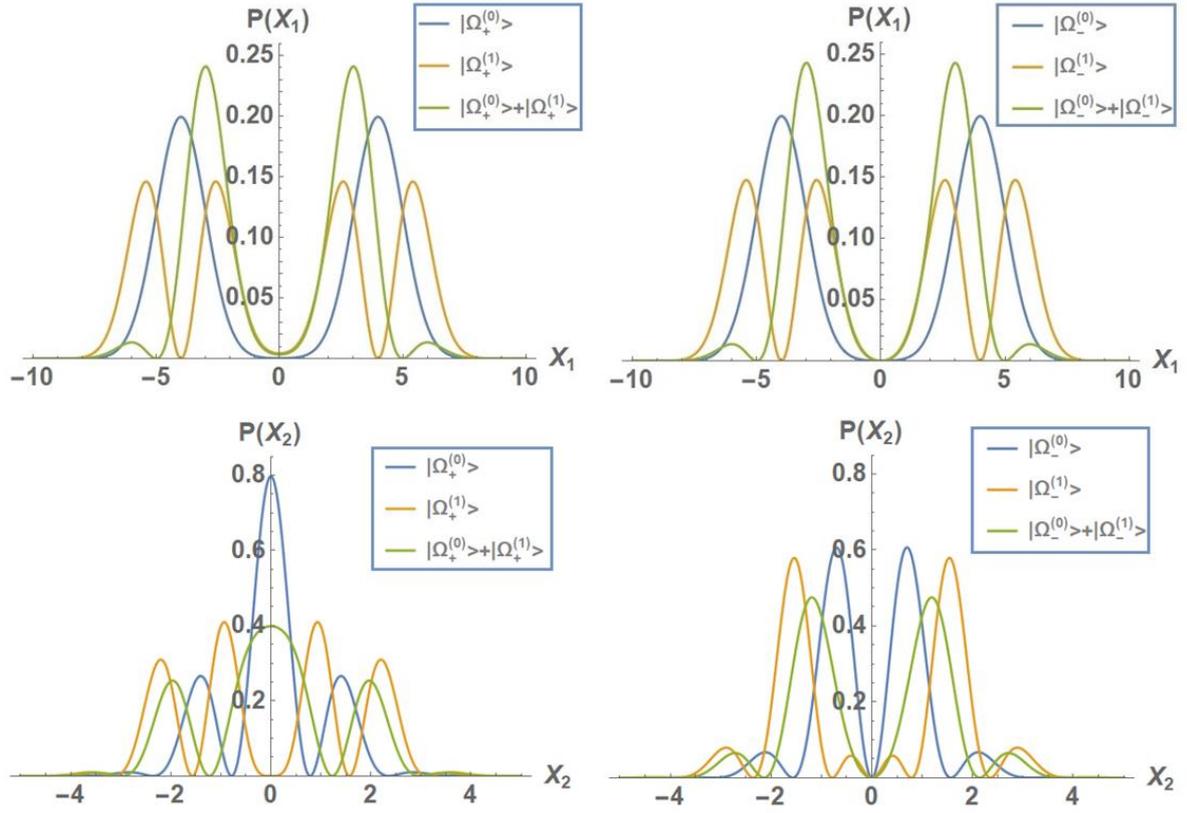

**Fig. 2.** Distributions of the quadrature component $P(X_1)$ (top two graphs) and $P(X_2)$ (bottom two plots) for the even states $|\Omega_+^{(0)}\rangle$, $|\Omega_+^{(1)}\rangle$, even superposition $|even\rangle = N_1^{(+)}\left(|\Omega_+^{(0)}\rangle + |\Omega_+^{(1)}\rangle\right)$ (left two graphics) and odd states $|\Omega_-^{(0)}\rangle$, $|\Omega_-^{(1)}\rangle$, $|odd\rangle = N_1^{(-)}\left(|\Omega_-^{(0)}\rangle + |\Omega_-^{(1)}\rangle\right)$ (right two graphics) for $\beta = 2$.

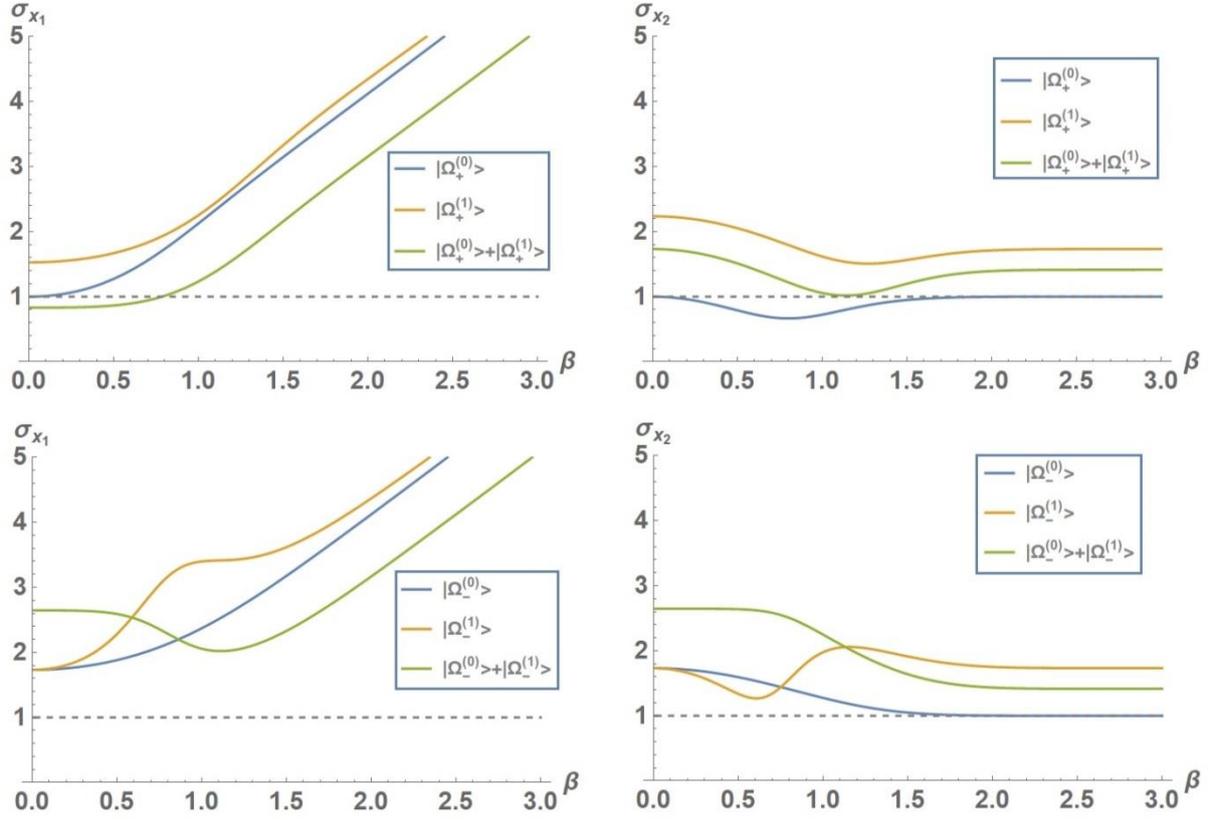

**FIG. 3.** Dependence of $\sigma_{x_1}$ (left two plots) and $\sigma_{x_2}$ (left two plots) on the size $\beta$ of the different SDlPSs: even states $|\Omega_+^{(0)}\rangle$, $|\Omega_+^{(1)}\rangle$ and $|even\rangle = N_1^{(+)}\left(|\Omega_+^{(0)}\rangle + |\Omega_+^{(1)}\rangle\right)$ (top two plots) and odd states $|\Omega_-^{(0)}\rangle$, $|\Omega_-^{(1)}\rangle$ and $|odd\rangle = N_1^{(-)}\left(|\Omega_-^{(0)}\rangle + |\Omega_-^{(1)}\rangle\right)$ (bottom two plots). The area, where either $\sigma_{x_1} < 1$ or $\sigma_{x_2} < 1$, corresponds to the squeezed properties of the states. Interestingly, superposition $|even\rangle$ possesses squeezing at small values of $\beta < 1$.

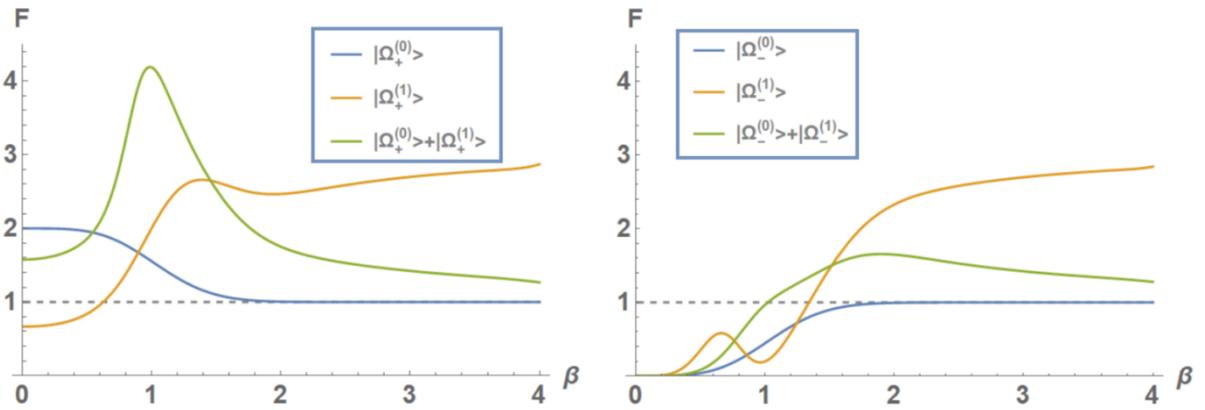

**Fig. 4.** Dependence of the factor Fano $F$ on the size $\beta$ of the different CV states under study: even states $|\Omega_+^{(0)}\rangle$, $|\Omega_+^{(1)}\rangle$ and $|even\rangle = N_1^{(+)}\left(|\Omega_+^{(0)}\rangle + |\Omega_+^{(1)}\rangle\right)$ (left two plots) and odd states $|\Omega_-^{(0)}\rangle$, $|\Omega_-^{(1)}\rangle$ and $|odd\rangle = N_1^{(-)}\left(|\Omega_-^{(0)}\rangle + |\Omega_-^{(1)}\rangle\right)$ (right two plots).

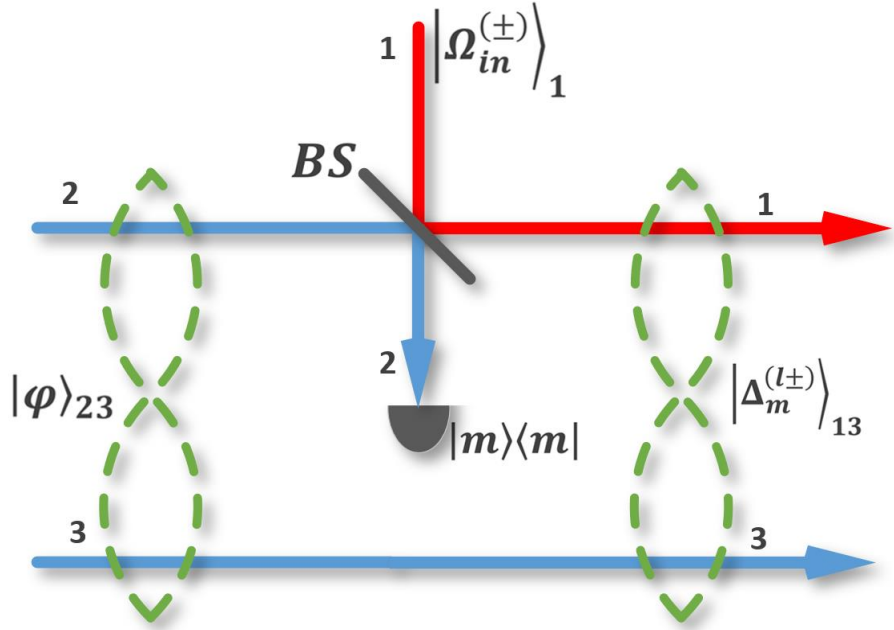

**FIG. 5.** Schematic representation for generating deterministic hybrid entanglement. The CV state, in general case $|\Omega_{in}^{(\pm)}\rangle$ in Eqs. (9,10), characterized by its parity is mixed with delocalized photon $|\varphi\rangle_{23}$ in Eq. (11) on the beam splitter (12) with arbitrary values of the transmittance and reflectance coefficients. Heralded entanglement $|\Delta_m^{(l\pm)}\rangle_{13}$ with some negativity either $\mathcal{N}_{2m}^{(l\pm)}$ or $\mathcal{N}_{2m+1}^{(l\pm)}$ occurs every time a measurement $m$ ($m$ can be either even or odd) is recorded in the second auxiliary mode. Under certain experimental conditions $(\beta, t)$, the entanglement can take on the maximum possible value $\mathcal{N}_{max} = 1$. Truncated versions of the SDlPSs in Eqs. (20,21) can also be used to generate the entangled states.

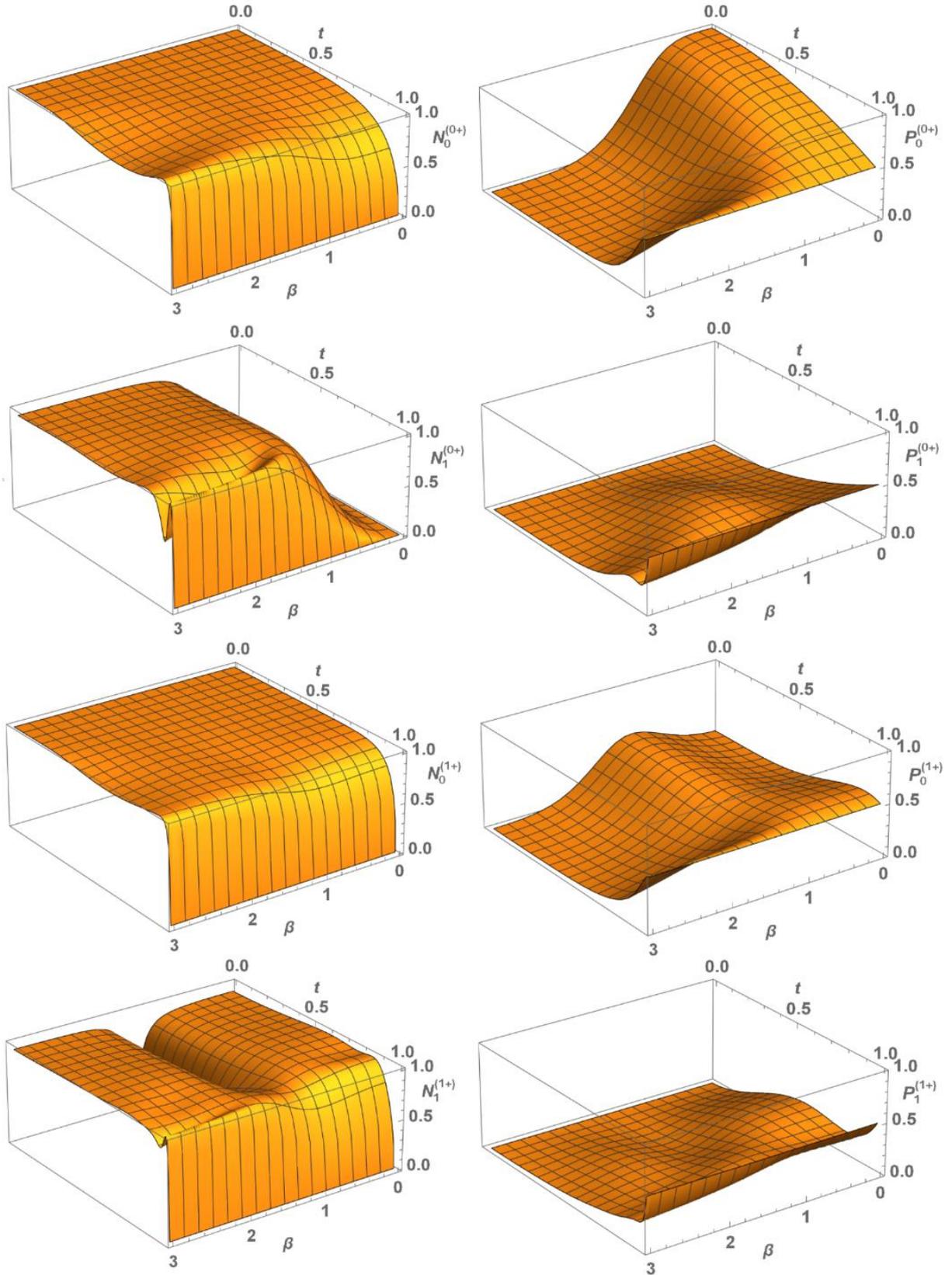

**FIG. 6.** Plots of the negativities $\left(\mathcal{N}_0^{(0+)}, \mathcal{N}_1^{(0+)}, \mathcal{N}_0^{(1+)}, \mathcal{N}_1^{(1+)}\right)$ (left side plots) and corresponding them success probabilities $\left(P_0^{(0+)}, P_1^{(0+)}, P_0^{(1+)}, P_1^{(1+)}\right)$ (right side plots) in dependency on $\beta$ and $t$.

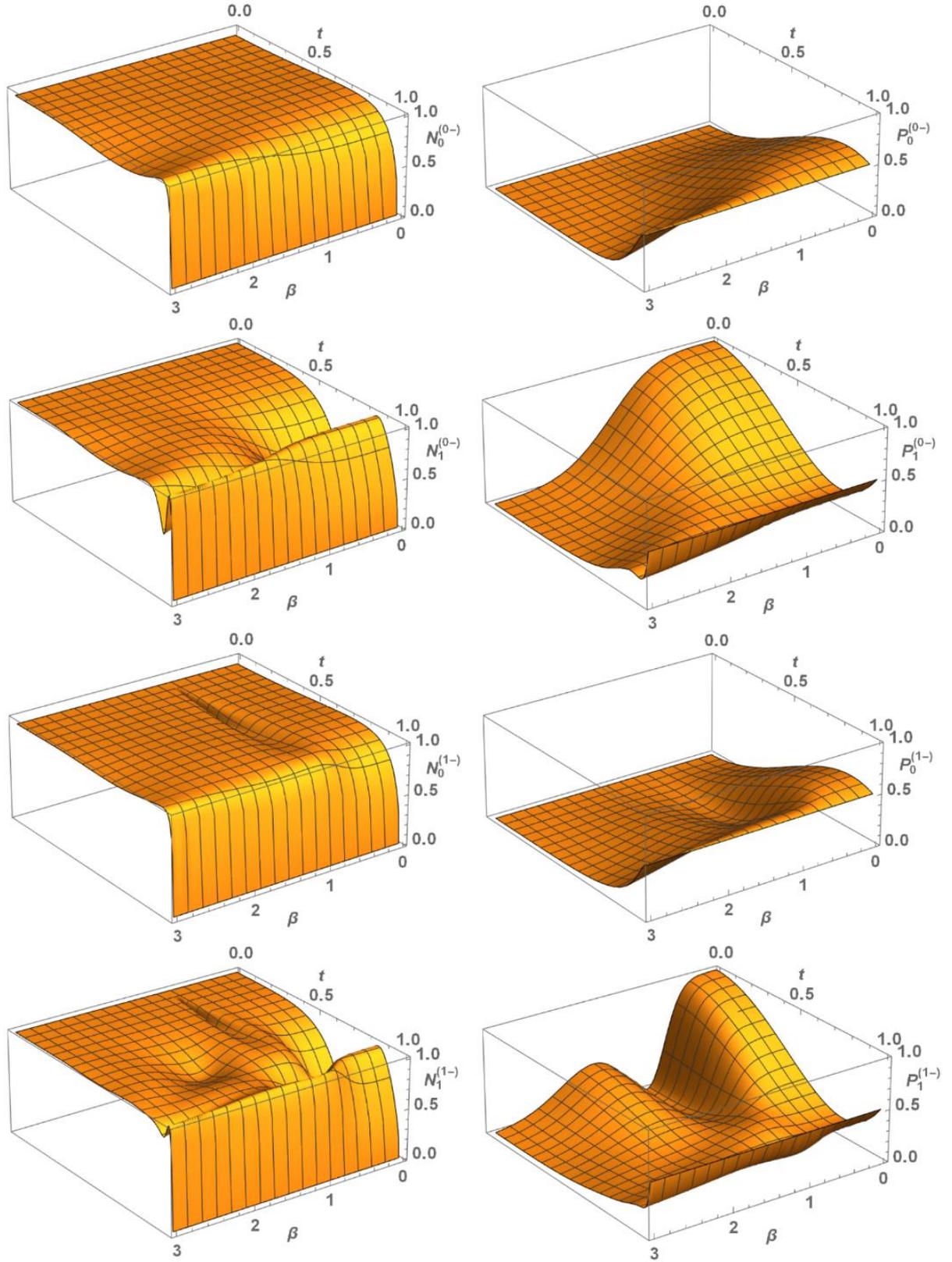

**FIG. 7.** Plots of the negativities $\left(\mathcal{N}_0^{(0-)}, \mathcal{N}_1^{(0-)}, \mathcal{N}_0^{(1-)}, \mathcal{N}_1^{(1-)}\right)$ (left side plots) and corresponding them success probabilities $\left(P_0^{(0-)}, P_1^{(0-)}, P_0^{(1-)}, P_1^{(1-)}\right)$ (right side plots) in dependency on $\beta$ and $t$.

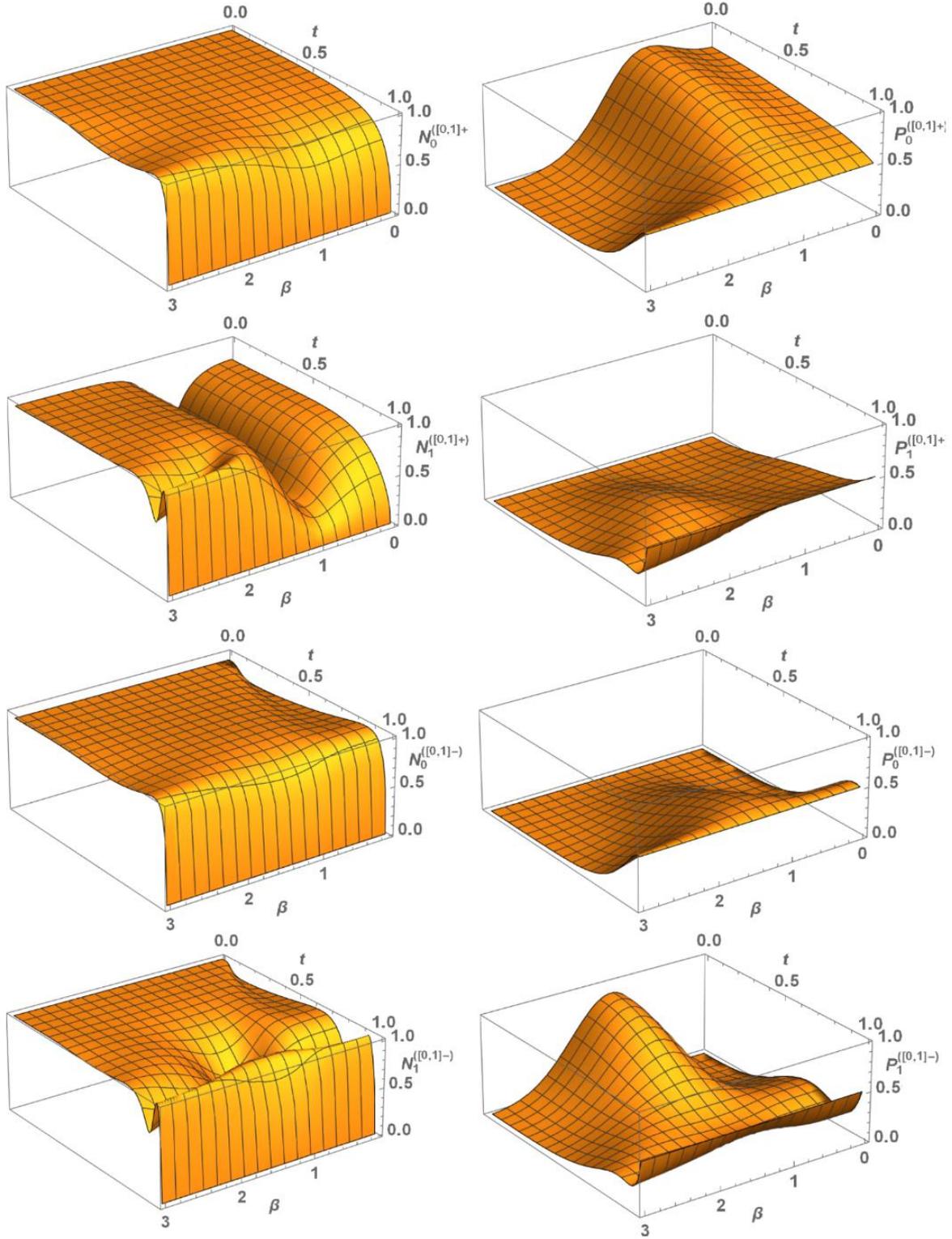

**FIG. 8.** Negativities $\left(\mathcal{N}_0^{([0,1]+)}, \mathcal{N}_1^{([0,1]+)}, \mathcal{N}_0^{([0,1]-)}, \mathcal{N}_1^{([0,1]-)}\right)$ (left side plots) of the conditional states generated from $|even\rangle \equiv |\Omega_{in}^{(+)}\rangle = N_1^{(+)}\left(|\Omega_+^{(0)}\rangle + |\Omega_+^{(1)}\rangle\right), |odd\rangle \equiv |\Omega_{in}^{(-)}\rangle = |odd\rangle = N_1^{(-)}\left(|\Omega_-^{(0)}\rangle + |\Omega_-^{(1)}\rangle\right)$ and corresponding them success probabilities $\left(P_0^{([0,1]+)}, P_1^{([0,1]+)}, P_0^{([0,1]-)}, P_1^{([0,1]-)}\right)$ (right side plots).